\documentclass{aa}
\usepackage{graphicx}
\usepackage{natbib}
\usepackage{txfonts}
\begin{document}

   \title{GK Boo and AE For: Two low-mass \\ eclipsing binaries with dwarf companions\thanks{This
   paper uses observations made at the South African Astronomical Observatory (SAAO)}}

   \titlerunning{GK Boo and AE For: Two low-mass eclipsing binaries with dwarf companions}

   \author{P. Zasche \inst{1}
            \and P. Svoboda\inst{2}
            \and R. Uhl\'a\v{r}\inst{3}}

   \authorrunning{P. Zasche et al.}

   \institute{Astronomical Institute, Faculty of Mathematics and Physics, Charles University Prague,
   CZ-180 00 Praha 8, V Hole\v{s}ovi\v{c}k\'ach 2, Czech Republic, \email{zasche@sirrah.troja.mff.cuni.cz}
   \and Private observatory, V\'ypustky 5, Brno, CZ-614 00 Czech Republic
   \and Private Observatory, Poho\v{r}\'{\i} 71, 25401 J\'{\i}lov\'e u Prahy, Czech Republic}

   \date{\today}

% \abstract{}{}{}{}{}
% 5 {} token are mandatory

  \abstract
  % context heading (optional)   % {} leave it empty if necessary
   {A study of late-type low-mass eclipsing binaries provides us with important information about the most
   common stars in the Universe.}
  % aims heading (mandatory)
   {We obtain the first light curves and perform period analyses of two neglected eclipsing binaries
   GK~Boo and AE~For to reveal their basic physical properties.}
  % methods heading (mandatory)
   {We performed both a period analysis of the times of the minima and a $BVR$ light curve analysis.
   Many new times of minima for both the systems were derived and collected from the data obtained by
   automatic and robotic telescopes. This allowed us to study the long-term period changes in these
   systems for the first time. From the light curve analysis, we derived the first rough estimates
   of the physical properties of these systems.}
  % results heading (mandatory)
   {We find that the analyzed systems are somewhat similar to each other. Both contain low-mass components
   of similar types, both are close to the Sun, both have short orbital period, and both contain another low-mass
   companions on longer orbits of a few years. In the case of GK~Boo, both components are probably of K3 spectral
   type, while the distant companion is probably a late M star. The light curve of GK~Boo is asymmetric, which
   probably causes the shift in the secondary minima in the $O-C$ diagram. System AE~For comprises two K7 stars,
   and the third body is a possible brown dwarf with a minimal mass of only about 47~$M_{Jup}$.}
  % conclusions heading (optional), leave it empty if necessary
   {We succeed in completing period and light curve analyses of both systems, although a more detailed
   spectroscopic analysis is needed to confirm the physical parameters of the components to a higher accuracy.}


  \maketitle

%________________________________________________________________

\section{Introduction}

Low-mass stars are the most common stars in our Galaxy (e.g. \citealt{2002Sci...295...82K}).
However, owing to their low luminosity, only these close to the Sun have been studied in detail and
many of them have never been analyzed. Hence, we focused on two rather neglected low-mass eclipsing
binary systems: GK Boo and AE For. Their light curves as well as their period modulation had never
been studied. Some studies indicate that most late-type stars are single (e.g.
\citealt{2006ApJ...640L..63L}), but the number of papers studying the multiplicity of the late-type
systems is still rather limited. Therefore, the incidence of multiples in late-type stars remains
unexplored.

The study of eclipsing binaries provide us with important information about the physical properties
of both of their components -- their radii, masses, and evolutionary status. However, when
considering only with the light curve, several assumptions have to be made. For the analysis
presented in this paper we also used the photometric data obtained by automatic and robotic
telescopes (such as ASAS, Pi of the sky, and SWASP). Thanks to these huge databases of
observations, the long-term evolution of these systems can be studied for the first time.

\section{GK Boo}

\subsection{Introduction}

The system GK Boo (= BD+37 2556, $V_{max} = 10.86$~mag) is an Algol-type eclipsing binary with an
orbital period of about 0.48~day. It is also a primary component of a visual double designated WDS
J14384+3632 in the Washington Double Star Catalog (WDS\footnote{http://ad.usno.navy.mil/wds/},
\citealt{WDS}). The secondary component of this double star is about 14$''$ distant, and is
probably gravitationally bound to GK~Boo itself. It is about 0.4~mag fainter, but since its
discovery in 1933 there has been no detectable mutual motion of the pair, hence the orbital period
is of about thousands of years (rough estimation from the Kepler's law).

The star is too faint, thus was not observed by Hipparcos satellite, and its distance is therefore
rather uncertain. \cite{2001KFNT...17..409K} introduced the parallax 30.29~mas, which is however
only an estimate. Its spectral type is also unknown, but the $B-V$ index derived from the Tycho
catalogue \citep{2000A&A...355L..27H}, $B-V=0.89$~mag indicates a spectral type of about K1. On the
other hand, the 2MASS infrared photometry \citep{2003yCat.2246....0C} gives $J-H = 0.527$~mag
(therefore a spectral type of K3). Finally, \cite{2006ApJ...638.1004A} introduced a temperature
corresponding to a spectral type of about K2-3. All these rough spectral estimates were taken from
\cite{1980ARA&A..18..115P} and \cite{2000asqu.book.....C}.

\subsection{Light curve}

The star was observed by the SuperWASP \citep{2006PASP..118.1407P} project and its complete light
curve (hereafter LC) is available. However, we did not use these data for the LC analysis because
these were not measured in any standard photometric filter. These data were only used to derive the
minima times (see below). We observed the target at the Ond\v{r}ejov observatory in the Czech
Republic with the 65-cm telescope equipped with the CCD camera. For the light curve analysis, only
the data from two nights in May 2011 were used (see the electronic data tables). The remaining
observations were used for the minima time derivation and to analyze the period changes in the
system (see below section \ref{GKBooPerAnal}). The observations were obtained in standard $B$, $V$,
and $R$ filters according to the specification of \cite{Bessell1990}.

At first, the complete LC was analyzed using the program {\sc PHOEBE} \citep{Prsa2005}, which is
based on the Wilson-Devinney algorithm (WD, \citealt{Wilson1971}). The derived quantities are as
follows: the secondary temperature $T_2$, the inclination $i$, the luminosities $L_i$, the gravity
darkening coefficients $g_i$, the albedo coefficients $A_i$, and the synchronicity parameters
$F_i$. The limb darkening was approximated using a linear law, and the values of $x_i$ were
interpolated from the van~Hamme's tables, given in \cite{vanHamme1993}.

At the beginning of the fitting process, we fixed the temperature of the primary component at
$T_1=4700$~K (corresponding to spectral type K3, \citealt{2000asqu.book.....C}). In the absence of
spectroscopy, the mass ratio was derived via a so-called "q-search method". This means that we
tried different values of mass ratio in the range 1.5 -- 0.5 in steps of 0.1 and tried to find the
best LC fit according to the lowest value of rms. Finally, we found that the best-fit solution was
reached with the value $q=M_2/M_1=0.9$, which agrees with both eclipses having almost equal depths.
For a given mass ratio, the semi-major axis was fixed to an appropriate value for the primary mass
to be equal to a typical mass of a particular spectral type (e.g. \citealt{1980ARA&A..18..115P},
\citealt{Hec1988}, or \citealt{1991A&ARv...3...91A}). With this approach, we were able to estimate
the masses, in addition to the radii of both components in absolute units.

\begin{figure}
  \centering
  \includegraphics[width=89mm]{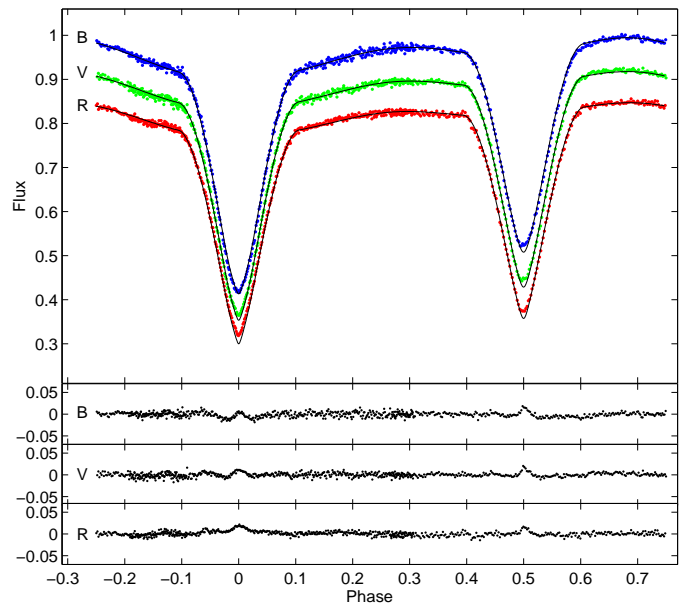}
  \caption{Light curves in $BVR$ filters for GK Boo, the solid lines represent the final fit. The residuals
  after the fit are plotted below. The curves are shifted along y-axis for reasons of clarity.}
  \label{GKBooLC}
\end{figure}

\begin{table}
 \caption{Light curve parameters of GK~Boo.}
 \label{LCparamGKBoo}
\small
 % \scriptsize
 %\tiny
 \centering
 \begin{tabular}{c c c}
 \hline\hline
 Parameter & \multicolumn{2}{c}{Value} \\
           & {\sc PHOEBE} & {\sc ROCHE} \\
 \hline
 $T_1$ [K]  & \multicolumn{2}{c}{4700$^*$} \\
 $T_2$ [K]  & 4540 $\pm$ 50         & 4615 $\pm$ 63 \\
 $q$ ($=\!{M_2}/{M_1}$) & 0.9$\pm$0.1 & 0.95 $\pm$ 0.12 \\
 $e$        & \multicolumn{2}{c}{0$^*$} \\
 $i$  [deg] & 89.83 $\pm$ 0.57      & 89.28 $\pm$ 0.37 \\
 $g_1$       & 0.00 $\pm$ 0.04      & 0.35 $\pm$ 0.05 \\
 $g_2$       & 0.00 $\pm$ 0.03      & 0.35 $\pm$ 0.05 \\
 $A_1$       & 0.00 $\pm$ 0.08      & 0.80 $\pm$ 0.05 \\
 $A_2$       & 1.00 $\pm$ 0.08      & 0.80 $\pm$ 0.05 \\
 $F_1$       & 1.892 $\pm$ 0.107    & 1.131 $\pm$ 0.096 \\
 $F_2$       & 1.866 $\pm$ 0.116    & 1.295 $\pm$ 0.108 \\
 $L_1$ (B) [\%] & 54.8 $\pm$ 1.9    & 52.4 $\pm$ 1.1 \\
 $L_2$ (B) [\%] & 45.2 $\pm$ 1.8    & 47.6 $\pm$ 1.0 \\
 $L_1$ (V) [\%] & 53.5 $\pm$ 1.5    & 51.6 $\pm$ 1.2 \\
 $L_2$ (V) [\%] & 46.5 $\pm$ 1.3    & 48.4 $\pm$ 1.1 \\
 $L_1$ (R) [\%] & 52.1 $\pm$ 1.4    & 51.1 $\pm$ 1.1 \\
 $L_2$ (R) [\%] & 47.9 $\pm$ 1.3    & 48.9 $\pm$ 1.0 \\
  \multicolumn{3}{c}{Spots:} \\
 l$_1$ [deg]   &         --        & 287.2 $\pm$ 7.9 \\
 b$_1$ [deg]   &         --        &  60.5 $\pm$ 3.2 \\
 r$_1$ [deg]   &         --        &  37.9 $\pm$ 2.0 \\
 k$_1$         &         --        &  0.75 $\pm$ 0.04 \\
 l$_2$ [deg]   &         --        &  63.3 $\pm$ 7.4 \\
 b$_2$ [deg]   &         --        &  47.4 $\pm$ 12.8 \\
 r$_2$ [deg]   &         --        &  28.7 $\pm$ 4.1 \\
 k$_2$         &         --        &  0.76 $\pm$ 0.04 \\ \hline %\cline{3-4}
 \multicolumn{3}{c}{Derived quantities:} \\
 $R_1$ [R$_\odot$] & 0.83 $\pm$ 0.18 & 0.89 $\pm$ 0.15 \\
 $R_2$ [R$_\odot$] & 0.86 $\pm$ 0.18 & 0.86 $\pm$ 0.14 \\              % $\sigma$ [mag] (B) &  0.0021 \\
 $M_1$ [M$_\odot$] & 0.73 $\pm$ 0.06 & 0.73 $\pm$ 0.06 \\              % $\sigma$ [mag] (V) &  0.0095 \\
 $M_1$ [M$_\odot$] & 0.66 $\pm$ 0.06 & 0.70 $\pm$ 0.06 \\\hline \hline % $\sigma$ [mag] (R) &  0.0226 \\
 Note: $^*$ - fixed.
 \end{tabular}
\end{table}

However, during the LC fitting process we found that the LC of GK~Boo is asymmetric. In particular,
the part of the LC near the secondary minimum is distorted in all $BVR$ filters. The brightness
just after the ascent from the secondary minimum (near the phase 0.6) is higher than the brightness
just before the descent (phase 0.4). The difference is about 0.022~mag in $B$, 0.018~mag in $V$,
and 0.017~mag in $R$ filter, respectively.

With the {\sc PHOEBE} code, we tried to fix the values of $A_i$ and $g_i$ to their appropriate
values of 0.5 and 0.32, respectively. However, after then we also allowed these parameters to be
fitted, because the fit is tighter (rms). However, probably owing to the asymmetry of the LC these
quantities converged to the rather improbable values given in Table \ref{LCparamGKBoo}, and the
shape of the observed LC could not be fitted properly. For the asymmetry of the curve, we also
tried to introduce a star spot on either of the components. However, no acceptable solution with
spot(s) was found to describe the shape of the light curve more accurately in the {\sc PHOEBE}
program. The parameters of the LC fit are given in Table \ref{LCparamGKBoo}, but these cannot
sufficiently describe the shape of the LC.

We therefore tried a different code, called {\sc ROCHE}, developed by Theo Pribulla
\citep{2004ASPC..318..117P}, which is also based on the WD code but has for instance also some
other computing methods and different controlling of the calculation process. With this program, we
used two star spots and similar input parameters as described above. At the beginning, the values
of $A_i$ and $g_i$ were fixed to the appropriate values of 0.5 and 0.32, respectively. However, to
achieve a tighter fit both $A_i$ and $g_i$ values were also varied across the range from 0 to 1 in
steps of 0.05 for both components. The synchronicity parameters $F_i$ converged to much more
reliable values. The value of mass ratio was fixed to $q=1.0$ and then also fitted as a free
parameter. This was possible because there is a clear distortion of the LC outside the minima (see
e.g. \citealt{Terrell2005}). For the fitting process, the two different limb darkening laws were
also tried, namely a linear and logarithmic. The latter one provides a much tighter fit to our
data. All of the resulting LC parameters are also given in Table \ref{LCparamGKBoo} (together with
parameters of two cooler spots located on the primary component -- longitude, latitude, radius and
temperature factor). As one can see, the two solutions clearly differ even outside their respective
error bars for some of the parameters.

The individual errors in the parameters were not taken from the WD code, but derived in the
following way. We computed a range of solutions for GK~Boo, which were then used for its error
estimation. All solutions with $\chi^2$ value close to the minimal one (5\% from our final
solution) were taken and the resultant values of parameters were used to compute the differences
between the parameters. The errors in the individual parameters were then computed as a maximum
difference and their individual WD errors, given by $max(a_i - a_{min}) + \delta a_i + \delta
a_{min}$.

This solution obtained with the {\sc ROCHE} program provides a much closer fit to the observed data
and is the fit plotted in Fig. \ref{GKBooLC}. The value of the eccentricity was fixed at 0 (for a
discussion about possible eccentricity see below). Our resultant parameters indicate that both the
components are still located on the main sequence, (as required because the age of the Universe
does not allow low-mass stars to have evolved from the main sequence). If we follow the assumption
of a K3V primary, then the secondary is also of K3V spectral type. These are consistent with the
photometric indices presented above, as well as with the individual masses and radii for these
types of stars (e.g. \citealt{Hec1988}). An undetectable value of the third light was also resulted
derived by this analysis. The presence of photospheric spots on both components of such a late
spectral type star is also foreseeable.

\subsection{Period analysis} \label{GKBooPerAnal}

To monitor the detailed long-term evolution of the system or its short-period modulation, we
collected all available published minima observations. Photometry from the SWASP
\citep{2006PASP..118.1407P}, ASAS \citep{2002AcA....52..397P}, and PiOfTheSky
\citep{2005NewA...10..409B} projects were used to derive many new minima times for GK Boo. All of
these data are given in Table \ref{MINIMA1}, which is available in electronic form only. The method
of \cite{Kwee} was used. Some of the data were of poor quality, but most were accurate enough to
perform a detailed period analysis of the system. The range of these data is about 12~years.

\begin{table}
 \caption{Final parameters of the long orbit for GK~Boo.}
 \label{FinalLITEGKBoo}
 \small
 \centering
 \begin{tabular}{c c c c}
 \hline\hline
 Parameter & Value \\
 \hline
 $HJD_0$       & 2454305.4570 $\pm$ 0.0006 \\
 $P$  [day]    & 0.47777174 $\pm$ 0.00000022 \\
 $p_3$ [day]   & 1472.7 $\pm$ 170.0 \\
 $p_3$ [yr]    & 4.032 $\pm$ 0.450 \\
 $A$  [day]    & 0.0126 $\pm$ 0.0012 \\
 $T_0$         & 2454263.3 $\pm$ 1108.3 \\
 $\omega_3$ [deg] & 56.54 $\pm$ 15.0 \\
 $e_3$          & 0.084 $\pm$ 0.267 \\
 $Q$ [$\cdot 10^{-10}$] & -1.071 $\pm$ 0.206   \\ \hline
 $f(M_3)$ [M$_\odot$] & 0.000633 $\pm$ 0.000002 \\
 $M_{3,min}$ [M$_\odot$] & 0.115 $\pm$ 0.001 \\
 $M_{3,60}$ [M$_\odot$]  & 0.134 $\pm$ 0.002 \\
 $M_{3,30}$ [M$_\odot$]  & 0.242 $\pm$ 0.005 \\
 $a_{12} \sin {i}$ [AU]  & 0.217 $\pm$ 0.108 \\
 $a_3$ [mas]  & 88.7 $\pm$ 9.8  \\
 \hline
 \end{tabular}
\end{table}

\begin{figure}
  \centering
  \includegraphics[width=89mm]{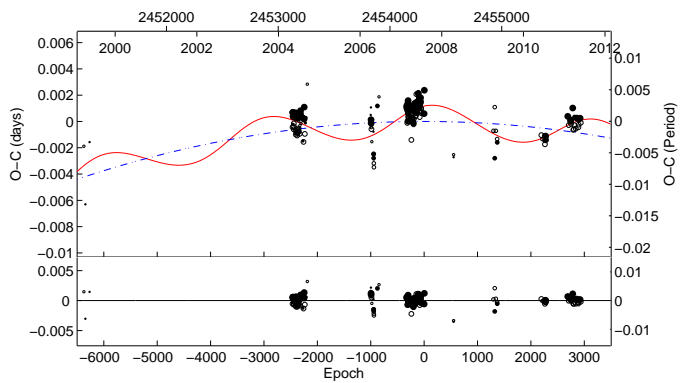}
  \caption{Periodic modulation of period GK Boo. Blue dashed line represents quadratic ephemeris,
  while red solid line stands for the LITE hypothesis. Residuals are plotted in the bottom plot.
  The larger the symbol, the higher the weight (higher the precision).}
  \label{GKBooLITE}
\end{figure}

We used these data to analyze the period modulation and found some interesting results. Applying
the hypothesis of a third body in the system (the so-called LIght-Time Effect, hereafter LITE,
described e.g. by \citealt{Irwin1959}), we found a weak period modulation with a period of about
four years. The final fit to the data together with the theoretical curve is shown in Fig.
\ref{GKBooLITE}. As one can see, there is also some long-term period evolution of the orbital
period (the blue dashed line), which was described as a quadratic term in ephemerides. It can be
understood as a slow period decrease caused by the mass loss from the system or mass flow between
the components (or even momentum loss, magnetic breaking, etc.). Another explanation is that this
is only part of the long-term period modulation, although we have only limited data coverage.

A more interesting finding is that of a period of about 4~years. Applying the LITE hypothesis, we
obtained a final set of parameters given in Table \ref{FinalLITEGKBoo}, namely the period of the
third body $p_3$, the semi-amplitude of the effect $A$, the time of periastron passage $T_0$, the
argument of periastron $\omega_3$, and the eccentricity $e_3$. Despite the low amplitude (about
only 1.8~minutes) of the LITE, most of the observed minima times are of higher precision and the
modulation is clearly visible. Table \ref{FinalLITEGKBoo} also provides the mass function of the
third body $f(M_3)$, which helps us to estimate its predicted mass.

\begin{figure}
  \centering
  \includegraphics[width=89mm]{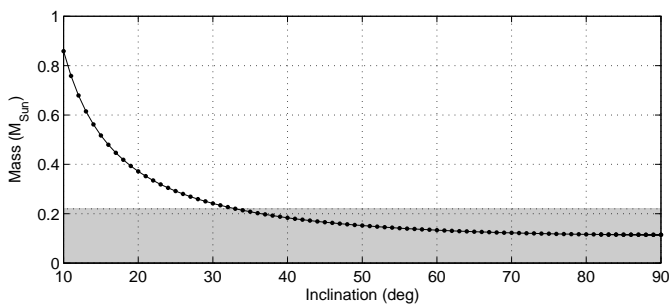}
  \caption{GK Boo: Mass of the third body based on from the LITE hypothesis with respect to the inclination
  between the orbits.}
  \label{GKBooMass}
\end{figure}

Having no information about the inclination between the orbits of the eclipsing pair and the
hypothetical third body, we plotted Fig. \ref{GKBooMass}, where a plot mass versus inclination is
shown. Assuming the coplanar orbits (i.e. $i_3=90^\circ \rightarrow M_3=M_{3,min}$), the resulted
minimum mass of the third body is only about 0.116~M$_\odot$, which places this body at the lower
end of stellar masses, hence we can rule out the hypothesis of a brown dwarf or even an exoplanet.
Despite of there being no upper limit to this mass (it goes to infinity with
$i_3\rightarrow0^\circ$), we can estimate a lower limit to the mass. Taking into account that no
third light is detected in the LC solution, e.g. $L_3/(L_1+L_2) < 0.01$ and assuming a
main-sequence star, we can estimate its mass to be lower than 0.22~M$_\odot$, which is shown in
Fig. \ref{GKBooMass} as a gray area. Further observations are still needed to confirm this
hypothesis with higher conclusiveness.

If we assume the parallax of GK Boo as given by \cite{2001KFNT...17..409K}, $\pi = 30.29$~mas, we
are also able to compute the angular distance of a hypothetical body to be about 89~mas. This
separation of components is well above the limit for modern stellar interferometers. However, there
is a problem with the brightness of the third component, which was found to be about more than five
magnitudes fainter than the eclipsing pair itself. With the brightness of about 11~mag for the
system, this makes a detection impossible. The magnitude difference of the third body with respect
to the close pair also clarify why no third light was detected in the LC solution.

\begin{figure}
  \centering
  \includegraphics[width=89mm]{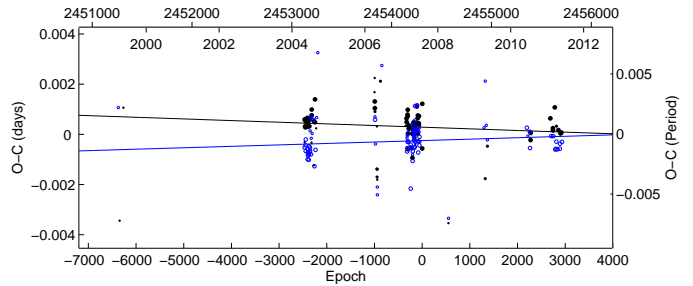}
  \caption{$O-C$ diagram of GK Boo with the apsidal motion hypothesis, black color is for primary minima,
  while blue one for secondary.}
  \label{GKBooAps}
\end{figure}

Another interesting result was a detection of displaced secondaries. This can be clearly seen in
more precise data points (SWASP and our new observations). That secondary minima occur at a
different phase of $\phi_2 \neq 0.5$ from the primary usually indicates that the system is on an
eccentric orbit. GK~Boo is a well-detached system, so the eccentric orbit cannot be ruled-out
easily. Therefore, we assumed an apsidal motion hypothesis for our data set of minima times. We
followed a procedure described by e.g. \cite{1983Ap&SS..92..203G} or \cite{1995Ap&SS.226...99G} and
obtained a set of apsidal motion parameters. The plot of residuals (after subtraction of the LITE
fit) with the apsidal motion fit is shown in Fig. \ref{GKBooAps}. It is obviously very slow because
the position of secondaries versus primaries changes only very slowly. The resultant values of
apsidal motion parameters are given in Table \ref{GKBooApsid}.

\begin{table}
 \caption{Apsidal motion parameters for GK~Boo.}
 \label{GKBooApsid}
 \small
 \centering
 \begin{tabular}{c c }
 \hline\hline
 Parameter & Value \\
 \hline
 $e$                       & 0.0944 $\pm$ 0.0068 \\
 $\omega$ [deg]            & 268.9 $\pm$ 2.5 \\
 $\dot \omega$ [deg/cycle] & 0.00026 $\pm$ 0.00001 \\
 $U$ [yr]                  & 1790 $\pm$ 50 \\
 \hline
 \end{tabular}
\end{table}

However, we have to rule out this hypothesis because it lead to unacceptable results. With some
information about the physical parameters of both components, we can use the apsidal motion
parameters to estimate the internal structure constant. The theoretical $\log$ k$_{2,{theor}}$
value taken from \cite{2004A&A...424..919C} should range from -1.35 to -1.65. However, the mean
value of $\log$ k$_2$ of both components that can be derived from our solution is very different,
even when k$_2 < 0$, which is unacceptable. Thus, the system is very probably on a circular orbit.

We may ask why the secondary minima deviate from the 0.5 phase. We published a finding that the
displaced secondary minima can also be present in contact binaries where no eccentric orbit is
possible \citep{Zasche2011}, so one cannot perform an apsidal motion analysis based only on the
minima times of a particular system. Some studies found that the secondary minimum is displaced
because of the distortion of the LC, thus any standard routine for deriving the time of minimum
(e.g. Kwee-van Woerden, bisector chord method or polynomial fitting) cannot be used properly
because these consider symmetric minima only. When using these methods to determine minima where
both ascending and descending branches have different slopes, we recover only a "false
eccentricity".

One can also ask about a significance of the fits presented in Figs. \ref{GKBooLITE} and
\ref{GKBooAps}. For this comparison, we summarized different approaches in Table
\ref{GKBooMethods}. In addition to the rms values, we also provide the values of BIC (Bayesian
Information Criterion, see e.g. \citealt{2007MNRAS.377L..74L}), which show the significance of the
fit. According to this method, the smaller the rms value, the tighter the fit. To conclude, our
final fit provides the smallest rms, but its significance is low and still highly speculative. This
is also caused by the poor data coverage, and large scatter in the minima and their low accuracy.
Determinations of new more precise minima are therefore needed to confirm or exclude this
hypothesis.

\begin{table}
 \caption{Methods of minima fitting for GK~Boo.}
 \label{GKBooMethods}
 \small
 \centering
 \begin{tabular}{r c c}
 \hline\hline
 Method of minima fitting & rms & BIC \\
 \hline
                             Linear ephemeris: & 0.00151 & 23.5 \\
                          Quadratic ephemeris: & 0.00119 & 29.2 \\
                    LITE and linear ephemeris: & 0.00080 & 51.0 \\
                 LITE and quadratic ephemeris: & 0.00074 & 56.4 \\
 LITE, quadratic ephemeris and apsidal motion: & 0.00060 & 72.6 \\
 \hline
 \end{tabular}
\end{table}

\section{AE For}

\subsection{Introduction}

The Algol-type system AE For (= HIP 14568, $V_{max} = 10.22$~mag) is also a poorly studied binary.
Its published spectral types range from K4 to M0, with the most probable one being K7V as derived
by \citep{2006A&A...460..695T}. The system was presented as a wide double with the star HD 19632
based on their similar parallaxes and proper motions (see \citealt{1994RMxAA..28...43P}).

Neither the light curve nor the radial velocity curve of AE~For have been studied. The star was
observed by the Hipparcos satellite and a few times also for the minima observations. It was also
continuously monitored with automatic photometric systems such as PiOfTheSky and ASAS. However, the
quality of these data do not allow us to use them for a LC analysis. The distance to the system was
derived from the Hipparcos data to be $d=31.5~$pc.

\subsection{Light curve}

We observed the star from the South African Astronomical Observatory (SAAO) in 2010, using the
classical one-channel photoelectric photometer mounted on the 50-cm telescope. All measurements
were carefully reduced to the Cousins E-region standard system \citep{Menzies} and corrected for
differential extinction.

Thanks to its orbital period close to one day, its complete light curve was observed once in
standard $BVR$ filters, with some overlapping points (about 170 data points in each filter were
obtained). Unfortunately, the quality of the data acquired for several nights was not very good,
hence the scatter in the curve is affected by these conditions. Two secondary and one primary
minima were observed (see below).

\begin{figure}
  \centering
  \includegraphics[width=89mm]{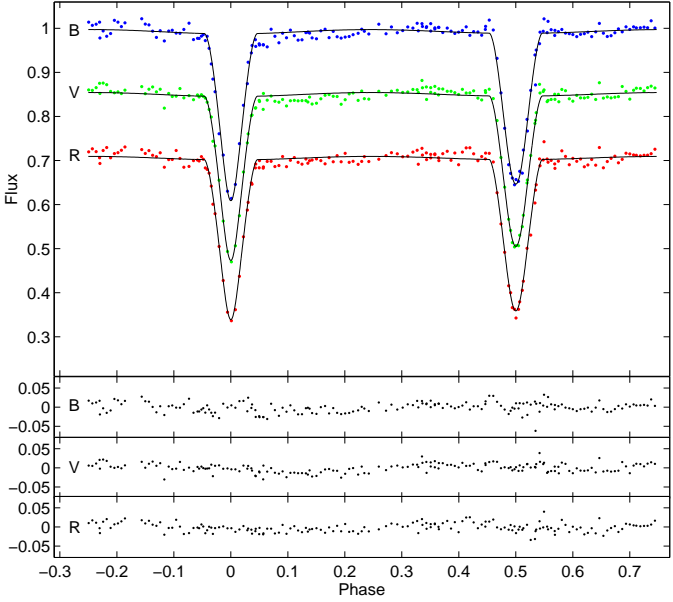}
  \caption{Light curve in $BVR$ filters for AE For, the solid line represents the final fit.
  The curves are shifted along the y-axis for greater clarity.}
  \label{AEForLC}
\end{figure}

We analyzed our data using the same computational procedure as for GK~Boo. The primary temperature
was fixed to the appropriate value of 4100~K (sp K7V), the eccentricity was fixed to 0, the values
of gravity darkening coefficients were fixed at 0.32, and the albedo coefficients to 0.5 (as
recommended for stars with convective envelopes), while the limb darkening coefficients were
interpolated from values given in \cite{vanHamme1993}. The computational approach was different for
the mass ratio $q$, which was fixed to $q=1.0$ because of the weak outside-eclipse ellipsoidal
variations and its detached configuration. In addition, the synchronicity parameters $F_i$ were set
to values of 1.0 for both components. The program {\sc ROCHE} was used and the resulting LC
parameters are given in Table \ref{LCparamAEFor}, while the final solution is presented in Fig.
\ref{AEForLC}.

\begin{table}
 \caption{Light curve parameters of AE~For.}
 \label{LCparamAEFor}
\small
 % \scriptsize
 %\tiny
 \centering
 \begin{tabular}{c c}
 \hline\hline
 Parameter  & Value \\
 \hline
 $T_1$ [K]   & {4100$^*$} \\
 $T_2$ [K]   &  4065 $\pm$ 48 \\
 $q$ ($=\!{M_2}/{M_1}$) &  1.0$^*$ \\
 $e$         & 0$^*$ \\
 $i$  [deg]  & 86.51 $\pm$ 0.31 \\
 $g_1 = g_2$ & 0.32$^*$ \\
 $A_1 = A_2$ & 0.50$^*$ \\
 $F_1 = F_2$ & 1.000$^*$ \\
 $L_1$ (B) [\%] & 63.2 $\pm$ 1.3 \\
 $L_2$ (B) [\%] & 36.8 $\pm$ 1.0 \\
 $L_1$ (V) [\%] & 63.1 $\pm$ 1.2 \\
 $L_2$ (V) [\%] & 36.9 $\pm$ 1.0 \\
 $L_1$ (R) [\%] & 62.6 $\pm$ 1.4 \\
 $L_2$ (R) [\%] & 37.4 $\pm$ 1.0 \\ \hline
 \multicolumn{2}{c}{Derived quantities:} \\
 $R_1$ [R$_\odot$] & 0.66 $\pm$ 0.10 \\
 $R_2$ [R$_\odot$] & 0.52 $\pm$ 0.08 \\
 $M_1$ [M$_\odot$] & 0.50 $\pm$ 0.05 \\
 $M_1$ [M$_\odot$] & 0.50 $\pm$ 0.05 \\\hline \hline
 Note: $^*$ - fixed.
 \end{tabular}
\end{table}

One can see that the secondary temperature $T_2$ is close to the value of $T_1$, indicating that
the components are similar. Thus, the estimated spectral types of both stars are probably K7V +
K7V. Both components are still located on the main sequence and their properties are in agreement
with the typical values of K7V stars (as presented by e.g. \citealt{Hec1988}). The third light was
also not detected here in any filter. In contrast to GK~Boo, the LC of AE~For seems to be
symmetric.

\subsection{Period analysis}

Similarly to GK~Boo, we tried to perform the period analysis of all available minima. The
collection of minima is much smaller, but thanks to the first observation by Hipparcos \citep{HIP}
these cover a longer time span than for GK~Boo. Several new minima were derived based on our new
observations from SAAO as well as those from the ASAS and PiOfTheSky surveys.

\begin{figure}
 \centering
 \includegraphics[width=89mm]{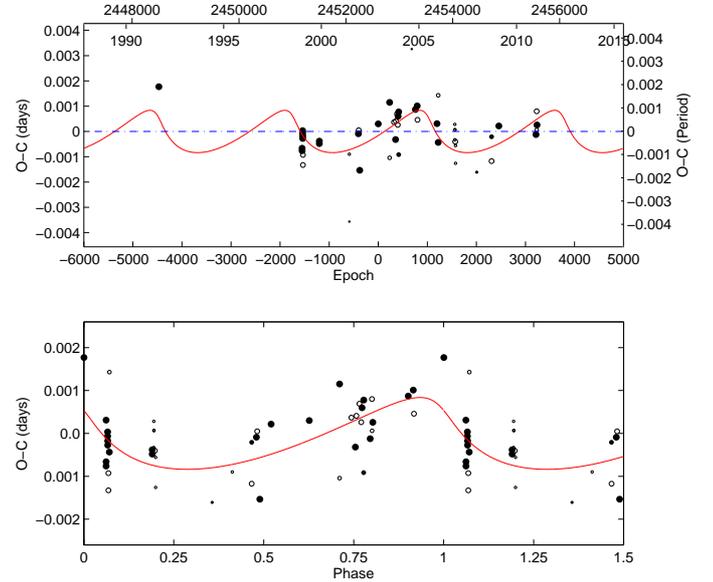}
 \caption{$O-C$ diagram of AE For. Up: With linear ephemeris. Bottom: With respect to the phase. The data
 points are fitted with the curve representing the third body hypothesis (see the text for details).}
 \label{AEForLITE}
\end{figure}

The same hypothesis as for GK~Boo was applied to the data points here. All of the minima times used
for the analysis are summarized in Table \ref{MINIMA2}, which is available in electronic form only.
As one can see from Fig. \ref{AEForLITE}, there is a clear variation in the minima times. We used
the same third-body hypothesis (LITE) as for GK~Boo, deriving a final fit to the data given by the
parameters in Table \ref{FinalLITEAEFor}. The LITE hypothesis resulted in a rather eccentric orbit,
although the result is affected by a relatively large error, hence maybe the $e_3$ value should be
lower. Only additional observations would help us confirm or refute this hypothesis, refine the
period, and possibly detect some long-term evolution of the period similar to that in GK~Boo,
because the first observation from Hipparcos deviates significantly from the fit. With the same
procedure as for GK~Boo, we computed the significance of the fits according to the BIC criterion
(see Table \ref{AEForMethods}). As one can see, the fit is still very poor and highly speculative.
However, using only the linear ephemeris, there remains a clear quasi-sinusoidal variation, which
needs some physical explanation.

From the LITE parameters, we were able to calculate the minimal mass of the third body (i.e.
coplanar orbits), which we found to be only about 47~$M_{Jup}$, which is even lower than the limit
of stellar masses. Therefore, if the orbits were coplanar (which only would be our assumption,
because the process of tidal coplanarization is very slow%e.g.1974Icar...23...51G
), the third body would very probably be a brown dwarf (exoplanets have masses about one half of
this value). With such a body, we reach minimal masses that can be detected by this method, because
the amplitude of LITE is comparable to the typical precision of individual minima-time
measurements. Whatever applies to the possible interferometric detection of GK~Boo companion also
applies here, because its luminosity is too low.

\begin{table}
 \caption{Final parameters of the long orbit for AE~For.}
 \label{FinalLITEAEFor}
 \small
 \centering
 \begin{tabular}{c c c c}
 \hline\hline
 Parameter & Value \\
 \hline
 $HJD_0$       & 2452605.97070 $\pm$ 0.00035 \\
 $P$  [day]    & 0.91820943 $\pm$ 0.00000012 \\
 $p_3$ [day]   & 2524.6 $\pm$ 149.6 \\
 $p_3$ [yr]    & 6.912 $\pm$ 0.409 \\
 $A$  [day]    & 0.00083 $\pm$ 0.00032 \\
 $T_0$         & 2453548.8 $\pm$ 413.1 \\
 $\omega_3$ [deg] & 146.2 $\pm$ 57.8 \\
 $e_3$          & 0.601 $\pm$ 0.414 \\ \hline
 $f(M_3)$ [M$_\odot$] & 0.000098 $\pm$ 0.000001 \\
 $M_{3,min}$ [M$_\odot$] & 0.047 $\pm$ 0.001 \\
 $M_{3,60}$ [M$_\odot$]  & 0.055 $\pm$ 0.001 \\
 $M_{3,30}$ [M$_\odot$]  & 0.098 $\pm$ 0.003 \\
 $a_{12} \sin {i}$ [AU]  & 0.167 $\pm$ 0.064 \\
 $a_3$ [mas]  & 117.2 $\pm$ 8.3  \\
 \hline
 \end{tabular}
\end{table}

\begin{table}
 \caption{Methods of minima fitting for AE~For.}
 \label{AEForMethods}
 \small
 \centering
 \begin{tabular}{r c c}
 \hline\hline
 Method of minima fitting & rms & BIC \\
  \hline
           Linear ephemeris: & 0.000255 & 24.4 \\
  LITE and linear ephemeris: & 0.000163 & 44.8 \\
  \hline
 \end{tabular}
\end{table}

\section{Discussion and conclusions}

We have derived preliminary light-curve solutions and period analyses of the poorly studied
Algol-type eclipsing binaries GK~Boo and AE~For, which we have found to have several interesting
and similar features. Since both of them are low-mass stars of very similar types (K3+K3 for
GK~Boo, and K7+K7 for AE~For), both of them have short orbital periods. Moreover, both are
relatively close to the Sun and also appear to contain third bodies in their systems, which cause a
periodic modulation of the orbital periods of both systems. Assuming a coplanar orbit, for AE~For
this third body appears to be a brown dwarf, which makes this system even more interesting.
However, more photometric and spectroscopic observations are needed to confirm or refute this
hypothesis.

The system GK~Boo has an asymmetric light curve, which is the probably accounts for the shift in
the secondary minimum in phase with the primary one. The apsidal motion hypothesis cannot explain
this discrepancy.

In general, if the third body hypothesis as proposed based on the period analysis is found to be
the correct one, here we have considered quite curious examples of hierarchical quadruple systems
of low masses. As far as we know, there are only a few similar multiple late-type systems for which
one of the components is an eclipsing binary (e.g. BB~Scl or MR~Del).

\begin{acknowledgements}

We thank the "ASAS", "SWASP" and "Pi of the sky" teams for making all of the observations easily
public available. PZ wish to thank the staff at SAAO for their warm hospitality and help with the
equipment. Authors are also grateful to Theo Pribulla for his helpful comments about the ROCHE
code. An anonymous referee is also acknowledged for his/her helpful and critical suggestions. This
work was supported by the Czech Science Foundation grant no. P209/10/0715 and also by the Research
Programme MSM0021620860 of the Czech Ministry of Education. This research has made use of the
SIMBAD database, operated at CDS, Strasbourg, France, and of NASA's Astrophysics Data System
Bibliographic Services.

\end{acknowledgements}

\begin{table}
 \caption{Heliocentric minima of GK Boo used for the analysis.}
 \label{MINIMA1}
 \scriptsize
 \centering
 \begin{tabular}{l l l l l l}
 \hline\hline
HJD - 2400000 & Error & Type & Filter & Reference\\
 \hline
 51260.8547  & 0.0007  & sec  & - & IBVS 5060 \\
 51273.9890  &         & prim & - & A.Paschke - Rotse \\
 51311.7377  & 0.0004  & prim & - & IBVS 5060 \\
 53128.46588 & 0.00024 & sec  & - & SWASP \\
 53128.70498 & 0.00041 & prim & - & SWASP \\
 53129.66141 & 0.00056 & prim & - & SWASP \\
 53130.61697 & 0.00047 & prim & - & SWASP \\
 53132.52777 & 0.00021 & prim & - & SWASP \\
 53135.39435 & 0.00010 & prim & - & SWASP \\
 53137.54360 & 0.00024 & sec  & - & SWASP \\
 53138.49891 & 0.00028 & sec  & - & SWASP \\
 53139.45491 & 0.00021 & sec  & - & SWASP \\
 53141.60560 & 0.00027 & prim & - & SWASP \\
 53152.59424 & 0.00025 & prim & - & SWASP \\
 53153.54982 & 0.00020 & prim & - & SWASP \\
 53154.50553 & 0.00017 & prim & - & SWASP \\
 53155.46071 & 0.00015 & prim & - & SWASP \\
 53156.41642 & 0.00014 & prim & - & SWASP \\
 53158.56492 & 0.00048 & sec  & - & SWASP \\
 53159.52074 & 0.00022 & sec  & - & SWASP \\
 53160.47640 & 0.00021 & sec  & - & SWASP \\
 53161.43169 & 0.00026 & sec  & - & SWASP \\
 53163.58313 & 0.00031 & prim & - & SWASP \\
 53164.53862 & 0.00031 & prim & - & SWASP \\
 53165.49392 & 0.00015 & prim & - & SWASP \\
 53166.44965 & 0.00020 & prim & - & SWASP \\
 53169.55361 & 0.00034 & sec  & - & SWASP \\
 53170.50939 & 0.00024 & sec  & - & SWASP \\
 53171.46489 & 0.00034 & sec  & - & SWASP \\
 53172.42077 & 0.00031 & sec  & - & SWASP \\
 53175.52707 & 0.00024 & prim & - & SWASP \\
 53176.48289 & 0.00047 & prim & - & SWASP \\
 53180.54262 & 0.00022 & sec  & - & SWASP \\
 53181.49835 & 0.00017 & sec  & - & SWASP \\
 53182.45376 & 0.00018 & sec  & - & SWASP \\
 53183.41009 & 0.00076 & sec  & - & SWASP \\
 53191.53267 & 0.00080 & sec  & - & SWASP \\
 53192.48678 & 0.00031 & sec  & - & SWASP \\
 53193.44381 & 0.00026 & sec  & - & SWASP \\
 53194.39850 & 0.00045 & sec  & - & SWASP \\
 53197.50383 & 0.00080 & prim & - & SWASP \\
 53198.46046 & 0.00023 & prim & - & SWASP \\
 53199.41623 & 0.00026 & prim & - & SWASP \\
 53203.47632 & 0.00029 & sec  & - & SWASP \\
 53209.44918 & 0.00047 & prim & - & SWASP \\
 53215.42110 & 0.00040 & sec  & - & SWASP \\
 53220.43588 & 0.00083 & prim & - & SWASP \\
 53221.39317 & 0.00039 & prim & - & SWASP \\
 53226.40796 & 0.00058 & sec  & - & SWASP \\
 53231.42630 & 0.00045 & prim & - & SWASP \\
 53232.38275 & 0.00032 & prim & - & SWASP \\
 53237.39732 & 0.00033 & sec  & - & SWASP \\
 53243.37030 & 0.00082 & prim & - & SWASP \\
 53248.38731 & 0.00078 & sec  & - & SWASP \\
 53259.85637 & 0.00078 & sec  & - & SWASP \\
 53827.68574 & 0.00085 & prim & - & SWASP \\
 53829.59740 & 0.00088 & prim & - & SWASP \\
 53830.55201 & 0.00069 & prim & - & SWASP \\
 53831.50729 & 0.00051 & prim & - & SWASP \\
 53832.46270 & 0.00015 & prim & - & SWASP \\
 53832.70138 & 0.00024 & sec  & - & SWASP \\
 53833.65681 & 0.00047 & sec  & - & SWASP \\
 53837.47803 & 0.00029 & sec  & - & SWASP \\
 53851.57305 & 0.00105 & prim & - & SWASP \\
 53852.52690 & 0.00049 & prim & - & SWASP \\
 53853.48213 & 0.00027 & prim & - & SWASP \\
 53854.43799 & 0.00032 & prim & - & SWASP \\
 53855.39312 & 0.00090 & prim & - & SWASP \\
 53855.63171 & 0.00037 & sec  & - & SWASP \\
 53856.58695 & 0.00051 & sec  & - & SWASP \\
 53887.40789 & 0.00024 & prim & - & SWASP \\
 53901.50285 & 0.00080 & sec  & - & SWASP \\
 54140.62657 & 0.00022 & prim & - & SWASP \\
 54149.70458 & 0.00021 & prim & - & SWASP \\
 54150.65998 & 0.00028 & prim & - & SWASP \\
 54153.76487 & 0.00119 & sec  & - & SWASP \\
 54154.71989 & 0.00026 & sec  & - & SWASP \\
 54155.67581 & 0.00061 & sec  & - & SWASP \\
 54156.63161 & 0.00053 & sec  & - & SWASP \\
 54157.58667 & 0.00065 & sec  & - & SWASP \\
 54159.73799 & 0.00019 & prim & - & SWASP \\
 54160.69347 & 0.00018 & prim & - & SWASP \\
 54161.64928 & 0.00022 & prim & - & SWASP \\
 54162.60325 & 0.00019 & prim & - & SWASP \\
 \hline \hline
\end{tabular}
\end{table}

\begin{table}
 \caption{Minima of GK Boo, cont.}
 \label{MINIMA1}
 \scriptsize
 \centering
 \begin{tabular}{l l l l l l}
 \hline\hline
HJD - 2400000 & Error & Type & Filter & Reference\\
 \hline
 54163.55942 & 0.00065 & prim & - & SWASP \\
 54165.70909 & 0.00037 & sec  & - & SWASP \\
 54166.66462 & 0.00022 & sec  & - & SWASP \\
 54167.62053 & 0.00034 & sec  & - & SWASP \\
 54170.72638 & 0.00023 & prim & - & SWASP \\
 54171.68178 & 0.00043 & prim & - & SWASP \\
 54189.59755 & 0.00032 & sec  & - & SWASP \\
 54190.55322 & 0.00019 & sec  & - & SWASP \\
 54191.50702 & 0.00053 & sec  & - & SWASP \\
 54194.61418 & 0.00029 & prim & - & SWASP \\
 54195.57050 & 0.00029 & prim & - & SWASP \\
 54202.49743 & 0.00046 & sec  & - & SWASP \\
 54206.55909 & 0.00039 & prim & - & SWASP \\
 54208.46922 & 0.00027 & prim & - & SWASP \\
 54208.70823 & 0.00143 & sec  & - & SWASP \\
 54210.61908 & 0.00030 & sec  & - & SWASP \\
 54212.53056 & 0.00050 & sec  & - & SWASP \\
 54213.48698 & 0.00049 & sec  & - & SWASP \\
 54214.44267 & 0.00029 & sec  & - & SWASP \\
 54214.68147 & 0.00039 & prim & - & SWASP \\
 54215.63620 & 0.00057 & prim & - & SWASP \\
 54216.59233 & 0.00019 & prim & - & SWASP \\
 54217.54805 & 0.00038 & prim & - & SWASP \\
 54218.50359 & 0.00021 & prim & - & SWASP \\
 54219.45928 & 0.00030 & prim & - & SWASP \\
 54220.41494 & 0.00069 & prim &BVR& PS    \\
 54220.41498 & 0.00029 & prim & - & SWASP \\
 54222.56423 & 0.00066 & sec  &BVR& PS    \\
 54222.56444 & 0.00028 & sec  & - & SWASP \\
 54223.51949 & 0.00169 & sec  &BVR& PS    \\
 54223.51994 & 0.00108 & sec  & - & SWASP \\
 54224.47565 & 0.00035 & sec  & - & SWASP \\
 54225.43110 & 0.00022 & sec  & - & SWASP \\
 54225.67010 & 0.00026 & prim & - & SWASP \\
 54226.38671 & 0.00034 & sec  & - & SWASP \\
 54226.62544 & 0.00036 & prim & - & SWASP \\
 54227.58111 & 0.00026 & prim & - & SWASP \\
 54228.53676 & 0.00031 & prim & - & SWASP \\
 54230.44800 & 0.00019 & prim & - & SWASP \\
 54231.40365 & 0.00019 & prim & - & SWASP \\
 54231.64168 & 0.00048 & sec  & - & SWASP \\
 54232.59790 & 0.00032 & sec  & - & SWASP \\
 54233.55304 & 0.00025 & sec  & - & SWASP \\
 54234.50899 & 0.00034 & sec  & - & SWASP \\
 54235.46550 & 0.00054 & sec  & - & SWASP \\
 54236.42047 & 0.00028 & sec  & - & SWASP \\
 54236.65842 & 0.00085 & prim & - & SWASP \\
 54237.37517 & 0.00137 & sec  &BVR& PS    \\
 54239.52553 & 0.00091 & prim &BVR& PS    \\
 54249.55879 & 0.00027 & prim & - & SWASP \\
 54250.51471 & 0.00031 & prim & - & SWASP \\
 54251.47023 & 0.00057 & prim & - & SWASP \\
 54252.42644 & 0.00028 & prim & - & SWASP \\
 54254.57647 & 0.00059 & sec  & - & SWASP \\
 54256.48657 & 0.00172 & sec  &BVR& PS    \\
 54256.48655 & 0.00039 & sec  & - & SWASP \\
 54257.44222 & 0.00025 & sec  & - & SWASP \\
 54261.50367 & 0.00031 & prim & - & SWASP \\
 54262.45860 & 0.00020 & prim & - & SWASP \\
 54263.41487 & 0.00037 & prim & - & SWASP \\
 54265.56380 & 0.00079 & sec  & - & SWASP \\
 54266.51892 & 0.00042 & sec  & - & SWASP \\
 54267.47491 & 0.00020 & sec  & - & SWASP \\
 54268.43084 & 0.00017 & sec  & - & SWASP \\
 54271.53663 & 0.00070 & prim & - & SWASP \\
 54272.49225 & 0.00033 & prim & - & SWASP \\
 54273.44808 & 0.00026 & prim & - & SWASP \\
 54276.55314 & 0.00095 & sec  & - & SWASP \\
 54277.50813 & 0.00065 & sec  & - & SWASP \\
 54278.46353 & 0.00031 & sec  & - & SWASP \\
 54279.41942 & 0.00021 & sec  & - & SWASP \\
 54305.45758 & 0.00157 & prim &BVR& PS    \\
 54307.37045 & 0.00141 & prim &BVR& PS    \\
 54568.70649 & 0.00072 & prim & - & Piofthesky \\
 54568.94557 & 0.00097 & sec  & - & Piofthesky \\
 54925.36514 & 0.0001  & sec  & R & OEJV 107 \\
 54937.3112  & 0.0020  & sec  & Ir& IBVS 5918 \\
 54937.5462  & 0.0005  & prim & Ir& IBVS 5918 \\
 54947.34260 & 0.0014  & sec  & R & OEJV 107 \\
 54958.8085  & 0.0007  & sec  & V & IBVS 5894 \\
 54959.52489 & 0.0001  & prim & R & OEJV 107 \\
 55354.40380 & 0.00148 & sec  &BVR& PS \\
 55364.4367  & 0.0001  & sec  &BVRI& IBVS 5965 \\
 55385.45891 & 0.00131 & sec  &BVR& PS \\
  \hline \hline
\end{tabular}
\end{table}

\begin{table}
 \caption{Minima of GK Boo, cont.}
 \label{MINIMA1}
 \scriptsize
 \centering
 \begin{tabular}{l l l l l l}
 \hline\hline
HJD - 2400000 & Error & Type & Filter & Reference\\
 \hline
 55386.41381 & 0.00104 & sec  &BVR& PS \\
 55391.43075 & 0.00097 & prim &BVR& PS \\
 55392.38658 & 0.00113 & prim &BVR& PS \\
 55590.66335 & 0.00036 & prim & - & RU \\
 55599.50146 & 0.00018 & sec  & - & RU \\
 55616.46275 & 0.00005 & prim & B & PZ \\
 55616.46266 & 0.00012 & prim & R & RU \\
 55619.56796 & 0.00010 & sec  & R & RU \\
 55634.61898 & 0.00016 & prim & R & RU \\
 55640.58947 & 0.00019 & sec  & R & RU \\
 55644.41169 & 0.00005 & sec  & B & PZ \\
 55650.62303 & 0.00014 & sec  & R & RU \\
 55651.34031 & 0.00113 & prim & B & PZ \\
 55662.56706 & 0.00013 & sec  & R & RU \\
 55671.40664 & 0.00007 & prim & VR& PZ \\
 55685.50021 & 0.00012 & sec  & R & RU \\
 55687.41150 & 0.00011 & sec  &BVR& PZ \\
 55692.42853 & 0.00004 & prim &BVR& PZ \\
 55700.55071 & 0.00004 & prim &BVR& PZ \\
 55707.47805 & 0.00008 & sec  &BVR& PZ \\
 \hline \hline
\end{tabular}
\end{table}

\begin{table}
 \caption{Heliocentric minima of AE For used for the analysis.}
 \label{MINIMA2}
 \scriptsize
 \centering
 \begin{tabular}{l l l l l l}
 \hline\hline
HJD - 2400000 & Error & Type & Filter & Reference\\
 \hline
 48500.6581  & 0.001  & prim & Hp & Hipparcos \\
 51180.9089  &        & prim & R  & VSOLJ 37 \\
 51180.9090  & 0.0002 & prim & R  & VSOLJ 47 \\
 51191.0099  &        & prim & R  & VSOLJ 37 \\
 51191.0100  & 0.0002 & prim & R  & VSOLJ 47 \\
 51191.9279  &        & prim & V  & VSOLJ 37 \\
 51191.9280  & 0.0001 & prim & V  & VSOLJ 47 \\
 51196.9770  & 0.0001 & sec  & R  & VSOLJ 47 \\
 51196.9774  &        & sec  & R  & VSOLJ 37 \\
 51504.11886 &        & prim & I  & VSOLJ 37 \\
 51504.1190  & 0.0002 & prim & I  & VSOLJ 47 \\
 52065.14178 & 0.00121& prim & V  & ASAS     \\
 52065.60355 & 0.0025 & sec  & V  & ASAS     \\
 52235.0140  &        & prim & I  & VSOLJ 39 \\
 52240.9825  &        & sec  & I  & VSOLJ 39 \\
 52258.8860  & 0.0002 & prim & R  & VSOLJ 39 \\
 52605.9710  &        & prim & I  & VSOLJ 40 \\
 52818.07823 & 0.00077& prim & V  & ASAS     \\
 52818.53514 & 0.00165& sec  & V  & ASAS     \\
 52901.1754  &        & sec  & I  & VSOLJ 42 \\
 52929.1801  &        & prim & I  & VSOLJ 42 \\
 52936.0674  &        & sec  & I  & VSOLJ 42 \\
 52957.1865  &        & sec  & V  & VSOLJ 42 \\
 52970.0410  &        & sec  & I  & VSOLJ 42 \\
 52976.0097  &        & prim & I  & VSOLJ 42 \\
 52987.0267  &        & prim & V  & VSOLJ 42 \\
 52987.9466  &        & prim & I  & VSOLJ 42 \\
 53300.1379  &        & prim & V  & VSOLJ 43 \\
 53335.0300  &        & prim & V  & VSOLJ 43 \\
 53340.0796  &        & sec  & V  & VSOLJ 43 \\
 53705.0677  & 0.0002 & prim & V  & VSOLJ 44 \\
 53728.02219 & 0.00093& prim & V  & ASAS     \\
 53728.48316 & 0.00194& sec  & V  & ASAS     \\
 54039.75501 & 0.00027& sec  & -  & PiOfTheSky\\
 54040.67301 & 0.00045& sec  & -  & PiOfTheSky\\
 54047.1000  & 0.0001 & sec  & V  & VSOLJ 45 \\
 54052.60840 & 0.00149& sec  & -  & PiOfTheSky\\
 54448.81542 & 0.00039& prim & -  & PiOfTheSky\\
 54726.11607 & 0.00187& prim & V  & ASAS      \\
 54726.57421 & 0.00131& sec  & V  & ASAS      \\
 54862.9297  & 0.0002 & prim & V  & VSOLJ 50  \\
 55558.0139  &        & prim & V  & VSOLJ 51  \\
 55570.41065 & 0.00072& sec  & BVR& PZ - SAAO \\
 55571.32812 & 0.00105& sec  & BVR& PZ - SAAO \\
 55576.37847 & 0.00065& prim & BVR& PZ - SAAO \\
  \hline \hline
 \end{tabular}
\end{table}

\end{document}